# A CLASSIFICATION SCHEME OF AMINO ACIDS IN THE GENETIC CODE BY GROUP THEORY*

**Sebastian Sachse,**

**Christian Roeder**

**Core Industry Solutions**
**SAP Banking**

**Financial Services,**
**CSC-Deutschland GmbH,**
**Abraham-Lincoln-Park 1,**
**65189 Wiesbaden, Germany**

ABSTRACT

We derive the amino acid assignment to one codon representation (typical 64-dimensional irreducible representation) of the basic classical Lie superalgebra osp(5|2) from biochemical arguments. We motivate the approach of mathematical symmetries to the classification of the building constituents of the biosphere by analogy of its success in particle physics and chemistry. The model enables to calculate polarity and molecular volume of amino acids to a good approximation.

## I. INTRODUCTION AND MOTIVATION

Mathematical symmetries play an important role in most models for explaining the organization and structure of dead matter which are scientifically accepted today. Here we would like to give a few examples only:

- The SU(3) – flavor model of Gell-Mann and Ne'eman[1] describes the multiplets of particles in high energy physics:

From this model the classification of hadrons has been derived. As an example consider the SU(3) hadron-octet representations with highest weight (1,1). This representation comprises the neutron and proton among other particles ($\Sigma^{\pm,0}$; $\Xi^{\pm}$; $\Lambda^0$ ) and as well the classification of the light mesons in the SU(3) meson-octet. All other particles built from the three quarks: up, down and strange and their anti-particles find their places in some other representation of SU(3). It should also be mentioned that this symmetry is not an exact property of the Hamiltonian operator since the strange quark exhibits a much greater mass than the up and down quark.

- The Grand Unified Theories (GUT) which explain in a unique form all interactions which play an important role in particle physics: the strong interaction and the electro-weak interaction.

These theories simultaneously explain the existence of the different gauge particles: the gluons for the strong interaction, the vector bosons for the weak interaction and the photon for the electromagnetic interaction. Furthermore, representations of the same symmetry group for including the Higgs scalars have been found, which account for the mass generation of the gauge particles. The most prominent examples of such GUTs in lower dimension are the SU(5) model of Georgi and Glashow[2] and the SO(10) model of Georgi and Fritzsch and Minkowski[3]. Both symmetries described above are internal symmetries of the Hamiltonian operator, which describes the interactions of the particles. They have no connection to the symmetries of 4-dimensional space-time.

- The classification of chemical elements in the periodic table, which has its justification by the atom model of Bohr is related to the representation theory of a symmetry group, i.e. to the representations of SO(4).

This symmetry has an outer part which is the SO(3) symmetry of spatial rotations. The other part is related to the Runge-Lenz vector and cannot be attributed to a *visible* symmetry nor can it be attributed to internal degrees of freedom of the Hamiltonian as those described in the first two examples. Therefore this symmetry is called a *hidden* symmetry.

It should be notified that as in the case of flavor symmetry a symmetry does not need to be reflected in a perfect manner. There might be some symmetry breaking in some few terms of the Hamiltonian operator (often in connection with a small multiplicative factor) which implies, that the symmetry is realized only approximately in the physical system. The symmetry breaking term might be switched on and off like an additional homogeneous magnetic field, which accounts for the **Zeeman effect** in the hydrogen atom and in a way lifts the degeneracy of the energy levels.

These three examples shall be sufficient in motivating the search for mathematical symmetries in the genetic code. It might give a feeling for the strength of symmetry principles in nature. It is difficult to conceive a physical world where such principles are not at hand. For the biosphere the discovery of such symmetries in its building blocks will have a revolutionary impact. Further information can be found in [4].

## II. BREAKING OF SYMMETRY AND THE STRUCTURE OF THE GENETIC CODE

The universal genetic code has been described in terms of an irreducible 64-dimensional representation (codon representation) of a basic classical Lie superalgebra. This model reproduces a branching process with the correct multiplet structure of the genetic table and serves as a model for the evolution of the genetic code in the early history of life.

The synthesis of proteins involves a process of decoding, where the genetic information stored in the DNA and carried by the mRNA (messenger RNA) is translated into amino acids through the use of tRNA (transfer RNA). Therefore the genetic code can be considered as a language which uses four different letters: A(denine), C(ytosine), G(uanine) and T(hymine) in DNA which is replaced by U(racil) in all types of RNA. From these letters three-letter words are formed: the codons. In this way one has 64 words to construct sentences called genes.

Each codon can be translated into one of twenty amino acids or a termination signal. This leads to a degeneracy of the code in the sense that different codons represent the same amino acid, that is, different words have the same meaning. Indeed, the codons which code for the same amino acids form multiplets as follows:

- 3 sextets Arg(R), Leu(L), Ser(S)
- 5 quadruplets Ala(A), Gly(G), Pro(P), Thr(T), Val(V)
- 2 triplets Ile(I), Term
- 9 doublets Asn(N), Asp(D), Cys(C), Gln(Q), Glu(E), His(H), Lys(K), Phe(F), Tyr(Y)
- 2 singlets Met(M), Trp(W)

The multiplet structure has resisted all explanations taken from first principles of biology, chemistry or physics. Therefore, an algebraic approach has been developed in order to find the underlying dynamical symmetry for the genetic code.

The basic concepts of algebraic models are the following[5].

a) Collective properties of the system as energy levels or transition rates between different states are described by operators, which are constructed in the universal enveloping algebra of the spectrum generating algebra (SGA). In the first algebraic model for the genetic code introduced by Hornos and Hornos[6,7,8] polarities of amino acids have been fitted with a few parameters.

b) The possible states of the system are associated to the vectors of an irreducible representation of the SGA. The Hamiltonian of the system is constructed as a linear combination of the Casimir operators of the SGA and its subalgebras. In order to guarantee that these Casimir operators can be diagonalized simultaneously the respective subalgebras must be taken from chains of subalgebras:

$$\mathbf{g} \supset \mathbf{g}_1 \supset \ldots \supset \mathbf{g}_k .$$

If the Casimir operators are taken from one single chain the system is said to exhibit a dynamical symmetry. It has to be emphasized that a general Hamiltonian not necessarily describes energy levels of the system. It might even have some undetermined signification, if it only reflects the degeneracy of the system.

The standard Hamiltonian of a dynamical symmetry may be modified by terms, which are of higher order in the Casimir elements of the last subalgebras (of the above mentioned chain) in order to achieve a fine tuning of the model. This has shown to be an essential feature for all models, which describe the degeneracy of the genetic code in terms of simple Lie groups and basic classical Lie Superalgebras and leads to the hypothesis which has been called *freezing* or *frozen accident*. It claims that the genetic code, evolving from simpler forms with lesser amino acids and therefore with a higher degeneracy to its present form could not complete the last branching step completely, i.e. for all the multiplets of the penultimate state: When a certain level of complexity had been achieved any further diversification had become lethal and therefore impossible.

The above-mentioned ideas have been investigated systematically within the context of typical codon representations of basic classical Lie superalgebras by Forger and Sachse[9,10]. The earlier work of Bashford et al.[11] based on the fundamental representation of $\mathbf{A(5|0)}$ does reproduce the structure of the so called *family boxes* i.e. the phenomenon that the first two nucleic bases of a codon determine the type of the amino acid up to an ambiguity rather than describing the multiplet structure of the genetic code exactly. Therefore it does not fit into the context of dynamical symmetries.

Admitting the possibility of *freezing* during the last step of the procedure, three schemes that do reproduce the degeneracy of the standard code have been found, all based on the codonrepresentation with highest weight (5/2,0,1) of the ortho-symplectic algebra $\mathbf{osp(5|2)}$ and differing only in the detailed form of the symmetry breaking pattern during the last step. The first three branching steps are given by the following chain of subalgebras:

$$\mathbf{osp(5|2)} \supset \mathbf{sp(2)} \oplus \mathbf{so(5)} \supset \mathbf{sl(2)} \oplus \mathbf{sl(2)} \oplus \mathbf{sl(2)} \supset \mathbf{sl(2)}_{12} \oplus \mathbf{sl(2)}.$$

The most natural scheme continues this process by a partial breaking of the second $\mathbf{sl(2)}$ into $\mathbf{o(2)}$ which respects the rules of freezing.

It allows a simple choice of Hamiltonian, in the sense used in [6] and explained in more detail in [4], namely the following

$$H_0 + \lambda_1 \, I_2\big(so(5)\big) + \lambda_2 \, I_4\big(so(5)\big) + \alpha_1 \, L_1^2 + \alpha_2 \, L_2^2 + \alpha_3 \, L_3^2 + \alpha_{12}(L_1 + L_2)^2 + \beta L_{3,z}^2 + \gamma_{12}((L_1 + L_2)^2 - 2)\,(L_{1,z} + L_{2,z})^2, \quad (1)$$

where the coefficients $\alpha_I$, $\beta$, $\lambda_i$ and $\gamma_{12}$ are free parameters and the other operators are the above mentioned Casimir operators. The symmetry perturbation i.e. the phenomenon described as freezing, stems from the last term of the Hamiltonian.

## III. AMINO ACID ASSIGNMENT

The principal remaining task is to perform the distribution of amino acids (aas) to the sub-representations occurring in each of the branching schemes mentioned above. It turns out that the first branching scheme not only exhibits the most simple Hamiltonian or freezing operator but also matches biological predictions to a large extent. Consequently, we want to restrict ourselves to explain the assignment of amino acids for this branching scheme (see tables 4+5). It has been taken from table 6 of Forger and Sachse[10].

Any branching scheme determines the distribution of amino acids up to a permutation of multiplets of equal dimension, which are

$$4 \times 6 \times 120 \times 9! \qquad .$$

This number cannot be reduced without further input. This input may come from the biochemical knowledge about the amino acids. In several papers it has been tried to mimic the evolution of the genetic code by the branching scheme. Even though this idea is attractive, since it would give a very intuitive meaning for the process of symmetry breaking, this is certainly not the only possible interpretation of the symmetry breaking. The subdivision of amino acids into different representations of subalgebras by the branching mechanism explained above could as well be derived from biochemical research on amino acids and not from the historical evolution of the genetic code which today is still under investigation. This is the approach we want to describe here.

We compare the grouping of amino acids given by our branching scheme with three different methods of grouping amino acids from purely biological and phenomenological facts.

1.) First the grouping of the physicochemical properties of amino acids should be taken into account. These properties have been described by Robert B. Russell and coworkers[12] and can be represented in a Venn diagram:

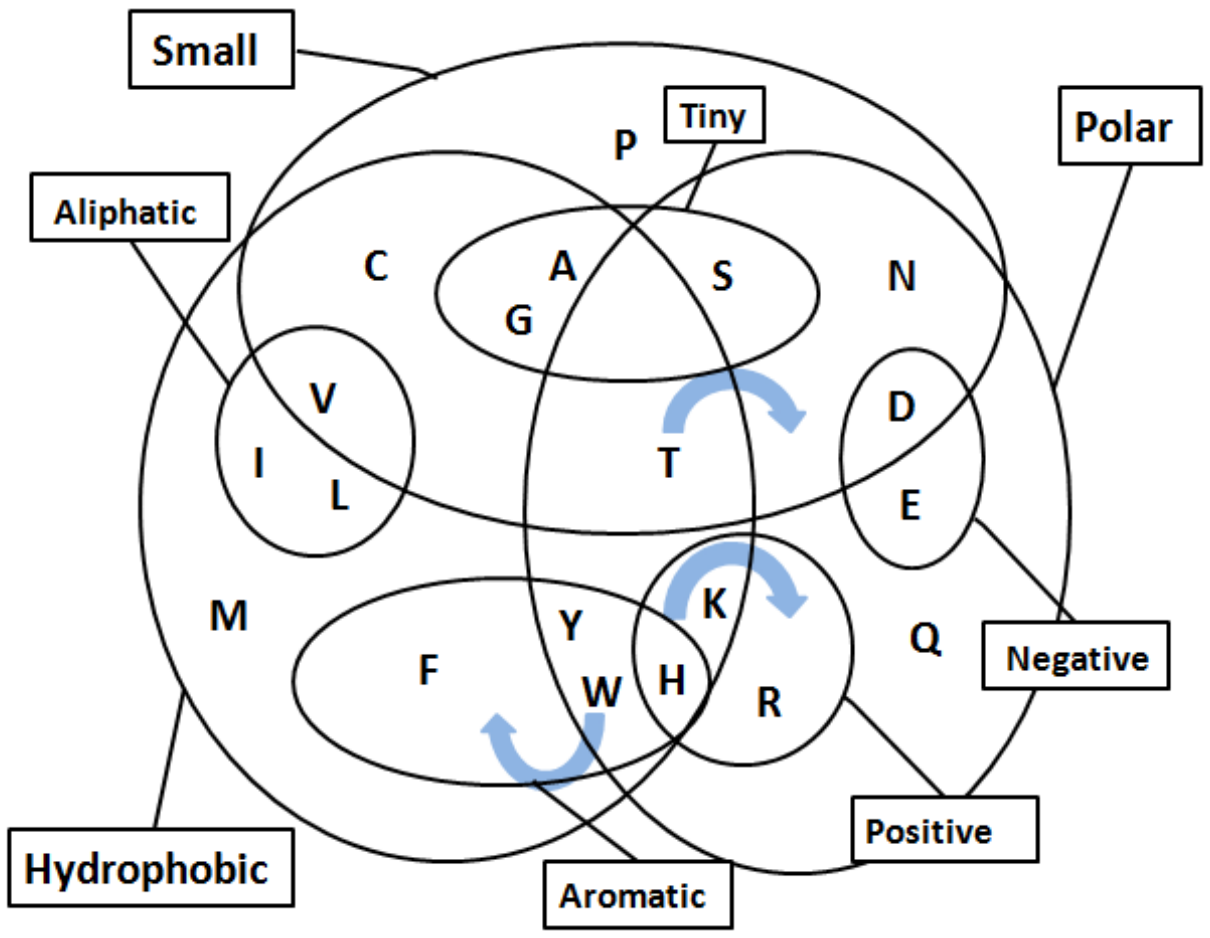

Picture 1: Venn diagram for amino acid properties, according to Russell[12]

The boundaries shown in this diagram are qualitative and not rigid. The properties of an amino acid depend not only on its chemical structure but may change with the environment in which it resides. Therefore some aas exhibit contradicting properties as hydrophobicity and polarity in the same diagram. Some other authors[13] do not accept that an amino acid at the same time would be hydrophobic and polar, considering only the dominant property. Those authors do not cite threonine and lysine as well as all charged amino acids within the hydrophobic group and cysteine not within the polar group[14]. Tryptophan which is the last amino acid with an ambiguity with respect to polarity and hydrophobicity, we rather consider in the hydrophobic subset, since it prefers to be buried in protein cores[12].

2.) Several research projects have reported groupings of amino acids by the mechanism which drives phylogenesis: genetic mutation or substitution rates of amino acids in coding genes of related species.

Substitution rates of aas are empirical data taken from the alignment of homologous gene sequences of proteins. (Gene sequences which share a common ancestor gene are called homologous.) Empirically determined substitution probabilities are a very important input in the calculation of phylogenetic relationships and trees[15,16,17]. To each choice of homologous gene samples individual substitution probabilities for each pair of amino acids are calculated. This gives rise to a matrix of substitution probabilities of dimension 20 x 20, which can be made symmetrical by multiplying the substitution probabilities with the frequencies of amino acid appearances in the sample. The symmetrization reflects the reversibility of the substitution process, when the sample has come to a stationary state of amino acid frequencies. From the notion of the symmetrized substitution rate matrix it is possible to perform the alignment of other related but not necessarily homologous genetic sequences. (It is certainly true, that there is some circular argument in this logic, which has to be carefully taken into account.) The probability of amino acid-substitution during protein evolution reflects the similarity of pairs of amino acids. This similarity is encoded therefore in the substitution rate matrices. It is however important to note, that there are various properties of amino acids – as given in picture 1 - and they may have different significance depending on the environment in which an amino acid resides. In some places a protein might be quite indifferent with respect to hydrophobicity in others this might be the preponderant factor for the choice of a suitable substituent. Therefore certain substitution rate matrices have proven to outperform others in the search for the correct alignments when working in special biological tissues and protein environments.

An algebraic method to determine groups from symmetrical substitution rate matrices has been discussed by Kosiol et al.[18] : The matrix has real eigenvalues which are smaller or equal to one. Calculating the corresponding eigenspaces Kosiol and colleagues have found so called *Almost Invariant Sets* of amino acids, which have the property of being conserved in most amino acid-replacements. They form almost invariant subspaces in the dynamics of protein evolution. The list of possible groupings for each number of possible subsets claimed by Kosiol is given in table 1 for two examples of substitution matrices: the PAM (point accepted mutations) and the WAG matrix.

{A G S P D E Q N H T K R M I L F Y V W}
{A G S P D E Q N H T K R M I L F Y V C} {W}
{A G S P D E Q N H T K R M I L F Y V} {W} {C}
{A G S P D E Q N H T K R M I V} {W} {Y F L} {C}
{A G S P D E Q N H T K R} {W} {Y F} {M I V L} {C}
{A G S P} {D E Q N H T K R V} {W} {Y F} {M I L} {V C}
{A G P} {D E Q N H} {T K R M I V} {W} {Y F} {L} {C S}
{A G} {D E Q N} {T K R M I V} {H Y} {W} {L} {F P} {C S}
{A G} {P} {D E Q N} {T K R M I} {H Y} {W} {F} {L} {V C S}
{A G} {P} {D E Q N} {T K R M} {H Y} {W} {F} {I} {L} {V C S}
{A G} {P} {D E Q N} {T K} {R I} {H} {Y} {W} {F} {M L} {V C S}
{F A S} {P} {G} {D E Q } {N L} {T K} {R} {H} {W} {Y} {I M} {V C}
{F A S} {P} {G} {D E Q} {N L} {T} {K} {R} {H} {W} {Y} {I M} {V C}
{F A} {P} {G} {T} {D E} {Q M} {N L} {K} {R} {H} {W} {Y} {I V} {C S}
{F A S} {P} {G} {T} {D E} {Q} {N L} {K} {R} {H} {W} {Y} {M} {I} {V C}
{F A} {P} {G} {S T} {D E} {Q} {N} {K} {R} {H} {W} {Y} {M} {L} {I} {V C}
{F A} {P} {G} {S T} {D E} {Q} {N} {K} {R} {H} {W} {Y} {M} {L} {I} {V} {C}
{F A} {P} {G} {S} {T} {D E} {Q} {N} {K} {R} {H} {W} {Y} {M} {L} {I} {V} {C}
{F A} {P} {G} {S} {T} {D} {E} {Q} {N} {K} {R} {H} {W} {Y} {M} {L} {I} {V} {C}
The 19 best groupings according to the PAM matrix

Table 1a): Result of the AIS algorithm for the PAM matrix, according to C.Kosiol[18]

{H R K Q S T A N E D P G C V I M L F Y W }
{H R K Q S T A N E D P G C V I M} { L F Y W}
{H R K Q S T A N E D P G C V I M} { L F Y} { W}
{H R K Q S T A N E D P G} {C V I} {M L F Y} { W}
{H R K Q S T A N E D P G} {C V} { I M L} { F Y} {W}
{H R K Q S T A N E D P} {G} {C V} {I M L} {F Y} {W}
{H R K Q S T A N E D} {G} {P} {C V} {I M L} {F Y} {W}
{H R K Q S T A} {N E D} {G} {P} {C V} {I M L} {F Y} {W}
{H R K Q} {N E D} { S T A G} {P} {C} {I V} {M L F} {Y} {W}
{R K H S A} {Q} {N E D} {G} {P} {C} {T I V} {M L F} {Y} {W}
{R K Q} {N G} {E D} {A S T} {P} {C} {I V} {H M L} {F} {Y} {W}
{R K Q} {E D} {N A S T} {G} {P} {C} {I V} {H} {M L} {F} {Y} {W}
{R K} {Q E} {D} {N G} {H A} {S T} {P} {C} {I V} {M L} {F} {Y} {W}
{R} {K} {Q E} {D} {N G} {H A} {S T} {P} {C} {I V} {M L} {F} {Y} {W}
{R} {K} {Q E} {D} {N G} {H A} {S T} {P} {C} {I V} {M} {L} {F} {Y} {W}
{R} {K} {Q} {E} {D} {N G} {H A} {S T} {P} {C} {I V} {M} {L} {F} {Y} {W}
{R} {K} {Q} {E} {D} {N G} {H A} {S} {T} {P} {C} {I V} {M} {L} {F} {Y} {W}
{R} {K} {Q} {E} {D} {N G} {H A} {S} {T} {P} {C} {I} {V} {M} {L} {F} {Y} {W}
{ R} {K} {Q} {E} {D} {N G} {H} {A} {S} {T} {P} {C} {I} {V} {M} {L} {F} {Y} {W}
b) The 19 best groupings for the WAG matrix

Table 1b): Result of the AIS algorithm for the WAG matrix, according to C.Kosiol[18]

Obviously the groupings differ for the two substitution matrices. Only very few preferences they have in common, for instance the subsets {C,V}; {D,E}; {F,Y}; {M,L}; {S,T}; in order to quote the groups which appear most frequently in both schemes - under the measure #(PAM) * #(WAG).
However this is not at all surprising and one explanation for the low consensus of predictions for invariant sets of amino acids has been given above: It depends on the sample of homologous genes which has led to the substitution matrix and certainly as well on the method with which the substitution rates have been derived from these probabilities.
We have started a similar investigation on the Blosum matrix[16]. Even though Blosum62 is not derived from a purely evolutionary (Markov-) process, the success of this matrix in protein sequence alignments makes it interesting to study almost invariant sets predicted by the same method. Therefore we made use of the program furnished by the Goldman group[19], in order to calculate the almost invariant sets as listed in table 2 on a symmetric version of the Blosum matrix.

| |
|---|
| ACDEFGHIKLMNPQRSTVYW |
| ACDEFGHIKLMNPQRSTVY W |
| ACFGILMSTV DEHKNPQRY W |
| ACFGILMSTV DEHKNQRY P W |
| ACILV DEGKNPQRST H FMY W |
| ACIL HNR FMTV DEGKPQS W Y |
| C ADEGKNQRST H FILMV P W Y |
| ACILV DEKNQRST FW G H M P Y |
| AGNST C DEKQR F H ILMV P W Y |
| C DENST F AG H ILMV KQR P W Y |
| C DENS F G H AILTV KQR M P W Y |
| C DES F G H AILTV KQR M N P W Y |
| C DS AEQT F G H ILV KR M N P W Y |
| AST C D F G H ILV M N P EQ KR W Y |
| C DS AE F G H ILV K M N PT Q R W Y |
| C D AES F G H ILV K M N P Q R T W Y |
| AS C D E F G H ILV K M N P Q R T W Y |
| C D E F G H K IL M N AP Q R S T V W Y |
| A C D E F G H IV K L M N P Q R S T W Y |
| A C D E F G H I K L M N P Q R S T V W Y |

Table 2: Result of the AIS algorithm for the Blosum matrix.

The invariant sets in this table repeat much lesser in different lines than in the two examples above. The only common feature, which we see in all of the three schemes, is the early separation of W, P and Y and perhaps a certain tendency of the ILMV – family to build an invariant set.
The list of equivalent methods for sequence comparison for amino acids is fairly long. An attempt to extract the common insight from eighteen of the most prominent alignment models has been made by May[20]. The author draws even a dendrogram to find a classification for these approaches. This however cannot alleviate the contradictions of groupings for amino acids derived from the different models. The context in which the substitutions matrices have been found in each model is not included in the resulting data but some additional information which is a given fact *a priori*. Is it possible to avoid such *a priori* boundary conditions which conceal universal features in the classification of amino acids? Such an approach has been made in the search for conserved substitution groups.

## 3.) Substitution groups

Substitution groups have been investigated by Wu and Brutlag[21] in an attempt to find universal patterns of amino acid substitutions. This is only possible when the context in which a substitution takes place is the relevant information itself. Indeed the strongest information given by this inquiry are conserved substitutions groups, i.e. subsets of amino acids which, when they build a vicinity in a protein, show some solidarity: all other amino acids pay a high separation score when they want to substitute into this alliance. (This does not mean, that different alliances of the same partner might not be allowed.) A conserved substitution group represents an invariant pattern in protein sequences and at the same time forms its proper context group. The method to find such substitution groups defers significantly from the methods to determine almost invariant sets described above: For each possible context (i.e. set of amino acids, in which each amino acid appears only once), the probability of substituting with one of twenty aas is its key ingredient. This implies a much bigger substitution matrix than in the former case. Instead of having the dimension 20 x 20, its size is $2^{20}$ x 20. Only the first 20 rows of this matrix form the kind of substitution matrix as in the search for almost invariant sets. Since the authors were interested in finding universal features, it was obvious that they had to use quite general databases of homologous proteins which do not select special environments. Furthermore they compared their inquiries on two databases which are gathered from quite different arguments: Blocks[16] and HSSP[22]. Whereas Blocks contains short and highly conserved regions of protein families, HSSP concentrates on longish protein sequences with the special purpose to take into account not only the 1-dimensional structure of a protein sequence derived from RNA but also its 3-dimensional folding. As result the authors give a classification of all empirically conserved amino acid substitution groups across both databases:

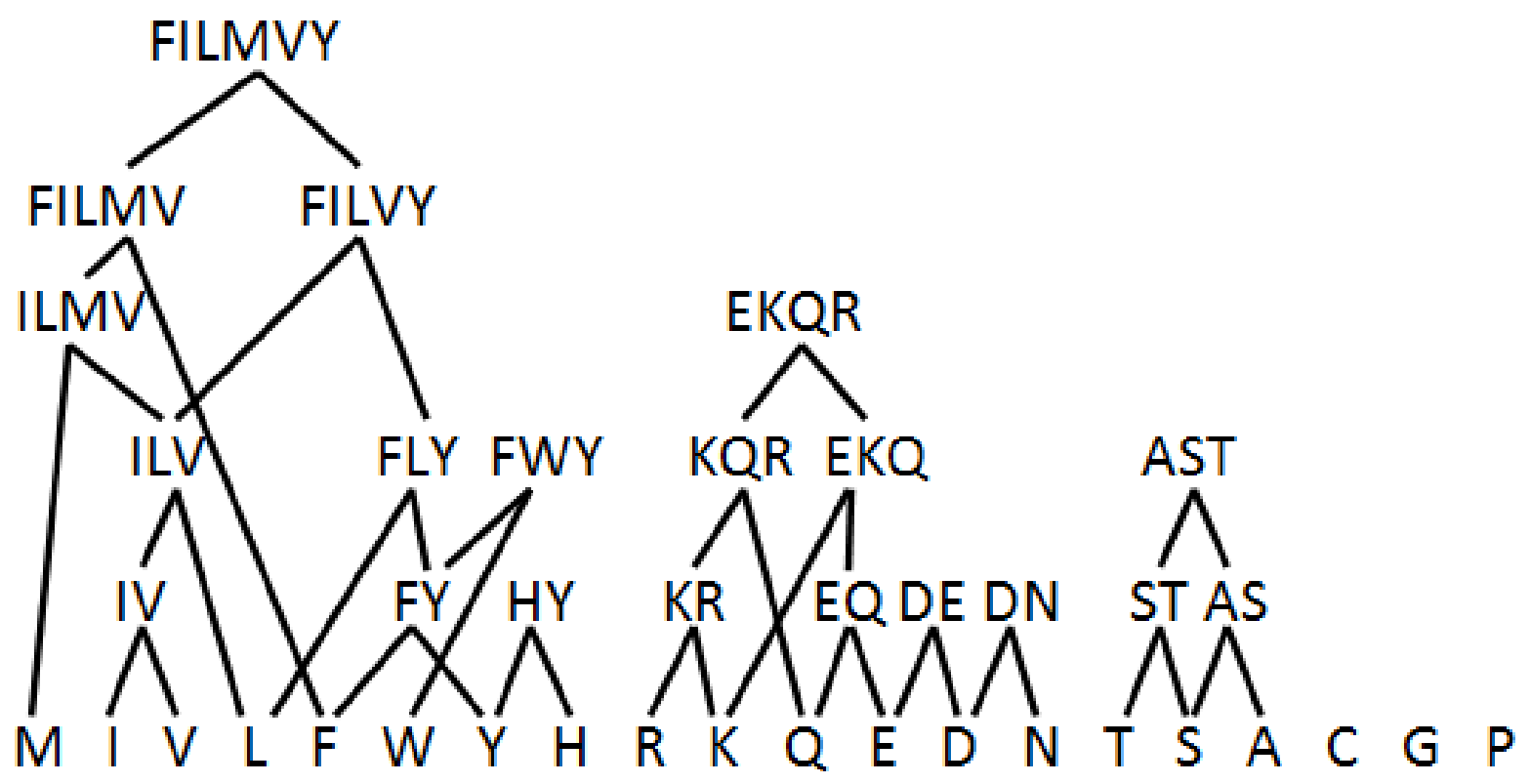

Table 3: Classification of all empirically conserved amino acid substitution groups. The substitution groups are linked by subsumption relationships. Cysteine, glycine and proline do not belong to any universal substitution group, according to Wu and Brutlag[21]

This classification scheme reflects many features, which are in good agreement with the physic-chemical properties of amino acids:

- hydrophobicity in the group FILMVY and its subgroups,
- aromatic side chains in the group FWY and HY,
- aliphatic behavior in the group ILV,
- polarity in the group EKQR and its derivates.

The only group which does not show any other common property except from size of the side chain is the substitution group formed by AST. Yet the high degree of congruence with the grouping in the Venn diagram remains remarkable.

Finally we have gathered sufficient material to justify our new classification scheme for the genetic code!

Picture 2 shows the weight diagram of the codon representation with highest weight (5/2,0,1) of osp(5|2). The weight vectors connected with green lines have multiplicity 2, those connected with red lines have multiplicity 3. Table 4 and 5 recalls the symmetry which reproduces the multiplet structure of the genetic code found in the previous work of M.Forger and S.Sachse**.** For the strategy of this investigation and the application of branching rules we refer to the cited articles[9,10]. Here we would like to emphasize, that the search among simple Lie Superalgebras and its codon representations for the possibility of explaining the structure of the genetic code has been complete. There are no other explanation patterns for the genetic code, which can be derived from this class of symmetries. In a recent publication F. Antoneli et al.[23] have shown, that the branching pattern derived for **osp (5|2)** together with the pattern based on **sp(6)**

investigated in the original work by Hornos & Hornos[6] are suitable for certain algebraic rules to find the codon assignment.

The branching scheme given in tables 4 and 5 provides the following information:

In the first row we state the chain of semi-simple Lie algebras and subalgebras which generates the correct multiplet structure for the codon representation upon its branching rules. The last subalgebras are of type **o(2)** $\subset$ **sl(2)** for the first and then for the second direct summand, indicated by its generator $L_z^2$.

Branching describes the splitting of irreducible representation of a Lie (super-)algebra into irreducible representations of a semi-simple subalgebra. The structure of the genetic code results after five branching steps. As mentioned before the last branching step is not performed by all multiplets: the sextets and the triplets of the penultimate step remain unbroken. This phenomenon finds an analogue in evolutionary theories of the genetic code proposed by biologists, who have denoted it as *freezing*. Further details can be found in [4]. We only would like to mention that it is an additional feature of our branching scheme that breaking the sextets and triplets in the last step would fit the multiplet structure perfectly into the family boxes of the well-known genetic table.

The first column of each branching step gives the highest weight of the resulting irreducible representation followed by its dimension in the second column.

| $\mathfrak{sp}(2) \oplus \mathfrak{so}(5)$ | | $\mathfrak{sl}(2) \oplus \mathfrak{sl}(2) \oplus \mathfrak{sl}(2)$ | | $\mathfrak{sl}(2)_{12} \oplus \mathfrak{sl}(2)$ | |
|---|---|---|---|---|---|
| Highest Weight | $d$ | Highest Weight | $d$ | Highest Weight | $d$ |
| $(1)-(1,1)$ | 32 | $(1)-(2)-(1)$ | 12 | $(3)-(1)$ | 8 |
| | | | | $(1)-(1)$ | 4 |
| | | $(1)-(1)-(2)$ | 12 | $(2)-(2)$ | 9 |
| | | | | $(0)-(2)$ | 3 |
| | | $(1)-(1)-(0)$ | 4 | $(2)-(0)$ | 3 |
| | | | | $(0)-(0)$ | 1 |
| | | $(1)-(0)-(1)$ | 4 | $(1)-(1)$ | 4 |
| $(0)-(0,3)$ | 20 | $(0)-(2)-(1)$ | 6 | $(2)-(1)$ | 6 |
| | | $(0)-(1)-(2)$ | 6 | $(1)-(2)$ | 6 |
| | | $(0)-(3)-(0)$ | 4 | $(3)-(0)$ | 4 |
| | | $(0)-(0)-(3)$ | 4 | $(0)-(3)$ | 4 |
| $(2)-(0,1)$ | 12 | $(2)-(1)-(0)$ | 6 | $(3)-(0)$ | 4 |
| | | | | $(1)-(0)$ | 2 |
| | | $(2)-(0)-(1)$ | 6 | $(2)-(1)$ | 6 |
| 3 subspaces | | 10 subspaces | | 14 subspaces | |

Table 4: Branching of the codon representation of **osp(5|2)** (first phase), according to Forger and Sachse[10]

| $\mathfrak{sl}(2)\oplus\mathfrak{sl}(2)\oplus\mathfrak{sl}(2)$ | | $\mathfrak{sl}(2)_{12}\oplus\mathfrak{sl}(2)$ | | $L^2_{3,z}$ | | $(L^2_{12,z}, L^2_{3,z})$ | |
|---|---|---|---|---|---|---|---|
| $2s_1-2s_2-2s_3$ | $d$ | $2s_{12}-2s_3$ | $d$ | $2s_{12}-2m_3$ | $d$ | $2m_{12}-2m_3$ | $d$ |
| $1-2-1$ | 12 | $3-1$ | 8 | $3-(\pm1)$ | 8 | $(\pm3)-(\pm1)$ | 4 |
| | | | | | | $(\pm1)-(\pm1)$ | 4 |
| | | $1-1$ | 4 | $1-(\pm1)$ | 4 | $(\pm1)-(\pm1)$ | 4 |
| $1-1-2$ | 12 | $2-2$ | 9 | $2-(\pm2)$ | 6 | $(\pm2)-(\pm2)$ | 4 |
| | | | | | | $0-(\pm2)$ | 2 |
| | | | | $2-0$ | 3 | $(\pm2)-0$ | 2 |
| | | | | | | $0-0$ | 1 |
| | | $0-2$ | 3 | $0-(\pm2)$ | 2 | $0-(\pm2)$ | 2 |
| | | | | $0-0$ | 1 | $0-0$ | 1 |
| $1-1-0$ | 4 | $2-0$ | 3 | $2-0$ | 3 | $(\pm2)-0$ | 2 |
| | | | | | | $0-0$ | 1 |
| | | $0-0$ | 1 | $0-0$ | 1 | $0-0$ | 1 |
| $1-0-1$ | 4 | $1-1$ | 4 | $1-(\pm1)$ | 4 | $(\pm1)-(\pm1)$ | 4 |
| $0-2-1$ | 6 | $2-1$ | 6 | $2-(\pm1)$ | 6 | $(\pm2)-(\pm1)$ | 4 |
| | | | | | | $0-(\pm1)$ | 2 |
| $0-1-2$ | 6 | $1-2$ | 6 | $1-(\pm2)$ | 4 | $(\pm1)-(\pm2)$ | 4 |
| | | | | $1-0$ | 2 | $(\pm1)-0$ | 2 |
| $0-3-0$ | 4 | $3-0$ | 4 | $3-0$ | 4 | $(\pm3)-0$ | 2 |
| | | | | | | $(\pm1)-0$ | 2 |
| $0-0-3$ | 4 | $0-3$ | 4 | $0-(\pm3)$ | 2 | $0-(\pm3)$ | 2 |
| | | | | $0-(\pm1)$ | 2 | $0-(\pm1)$ | 2 |
| $2-1-0$ | 6 | $3-0$ | 4 | $3-0$ | 4 | $(\pm3)-0$ | 2 |
| | | | | | | $(\pm1)-0$ | 2 |
| | | $1-0$ | 2 | $1-0$ | 2 | $(\pm1)-0$ | 2 |
| $2-0-1$ | 6 | $2-1$ | 6 | $2-(\pm1)$ | 6 | $(\pm2)-(\pm1)$ | 4 |
| | | | | | | $0-(\pm1)$ | 2 |
| 10 subspaces | | 14 subspaces | | 18 subspaces | | 26 subspaces | |

Table 5: Branching of the codon representation of **osp(5|2)** (second phase):
First option, according to Forger and Sachse[10]

Finally we assign the amino acids of the genetic code into table 5, such that we obtain the best agreement with the groupings described in chapters III.1. - III.3. This has been done by a systematical process – regarding physico-chemical properties as the driving force for a reliable grouping. At first we subdivide the amino acids into its hydrophobic and its polar subgroups. We take out from the hydrophobic group all amino acids which exhibit aromatic side chains. Among those we separate tryptophan since it shows the littlest susceptibility to invariant sets as shown in tables 1-2. It is put into one subrepresentation with the stop signal, which of course has no amino acid property but plays the role of the zero element in the group. It shows very low clustering in invariant sets as well and has been disregarded in the analysis of Kosiol even on the level of codons. The next amino acid unlikely to show up in invariant sets of tables 1, 2 and 3 is proline and thus gains as well a place in an early separating representation. Next we join pairs of doublets which have some structural similarity as there are the pairs of the glutamine and of the asparagine family. Additionally we have two amino acids which incorporate a sulfur atom: cysteine and methionine. Therefore we put them into some subrepresentation as well. Continuing this process we form amino acid families (AAF) indicated by xF in table 6. Putting amino acid families with the same physico-chemical properties into the biggest non occupied subrepresentations completes the process of amino acid assignment. There remain only very few ambiguities as for instance the interchange of serine and arginine or the interchange of the asparagine and the glutamine family.

| $\mathfrak{sl}(2) \oplus \mathfrak{sl}(2) \oplus \mathfrak{sl}(2)$ | | $\mathfrak{sl}(2)_{12} \oplus \mathfrak{sl}(2)$ | | $L^2_{3,z}$ | | $(L^2_{12,z}, L^2_{3,z})$ | |
|---|---|---|---|---|---|---|---|
| $2s_1 - 2s_2 - 2s_3$ | $d$ | $AAF$ | $d$ | $AAF$ | $d$ | $AA^{mol}$ | $d$ |
| $1-2-1$ Ala$^F$ ★ | 12 | Ala$^F$ | 8 | Ala$^F$ | 8 | Val$^{99}$ | 4 |
| | | | | | | Ala$^{71}$ | 4 |
| | | Gly | 4 | Gly | 4 | Gly$^{57}$ | 4 |
| $1-1-2$ Leu$^F$ ★ | 12 | Leu$^F$ | 9 | Leu$^{113}$ | 6 | | 4 |
| | | | | | | | 2 |
| | | | | Ile$^{113}$ | 3 | | 2 |
| | | | | | | | 1 |
| | | Cys$^F$ S | 3 | Cys | 2 | Cys$^{103}$ | 2 |
| | | | | Met | 1 | Met$^{131}$ | 1 |
| $1-1-0$ Ter$^F$ a ★ | 4 | Ter | 3 | Ter | 3 | | 2 |
| | | | | | | | 1 |
| | | Trp | 1 | Trp | 1 | Trp$^{186}$ | 1 |
| $1-0-1$ ★ | 4 | Pro | 4 | Pro | 4 | Pro$^{97}$ | 4 |
| $0-2-1$ • | 6 | Arg | 6 | Arg$^{+;156}$ | 6 | | 4 |
| | | | | | | | 2 |
| $0-1-2$ • | 6 | Thr$^F$ | 6 | Thr | 4 | Thr$^{101}$ | 4 |
| | | | | Lys | 2 | Lys$^{+;129}$ | 2 |
| $0-3-0$ • | 4 | Glu$^F$ | 4 | Glu | 4 | Gln$^{128}$ | 2 |
| | | | | | | Glu$^{-;129}$ | 2 |
| $0-0-3$ • | 4 | Asp$^F$ | 4 | Asp | 2 | Asn$^{115}$ | 2 |
| | | | | Asn | 2 | Asp$^{-;114}$ | 2 |
| $2-1-0$ a ★ | 6 | Phe$^F$ | 4 | Phe | 4 | Phe$^{147}$ | 2 |
| | | | | | | Tyr$^{163}$ | 2 |
| | | His | 2 | His | 2 | His$^{+;137}$ | 2 |
| $2-0-1$ (•) | 6 | Ser | 6 | Ser$^{87}$ | 6 | | 4 |
| | | | | | | | 2 |
| 10 subspaces | | 14 subspaces | | 18 subspaces | | 26 subspqaces | |

Table 6: Assignment of aas to the codon representation of osp(5 | 2)
**a** : aromatic side chain, ★: hydrophobic, • : polar,
**S** : sulfur, ± : charged.

Table 6 shows the final assignment of amino acids which we have found. The first row again gives the sequence of subalgebras during the last branching steps. The physico-chemical properties of the amino acids are indicated by the symbols explained in the legend of the table. In the final multiplet of each amino acid we indicate its molecular mass showing that we find some interesting mass multiplets as well. Molecular mass is of course only an approximate symmetry as in the case of the flavor symmetry. It is remarkable that we obtain a classification scheme were the property of hydrophobicity can be derived from the adherence of an amino acid to one of two representations only (i.e. the (1) – (1,1) representation of **sp(2)** ⊕ **so(5)** and (2 – 1 – 0) of **sl(2)** ⊕ **sl(2)** ⊕ **sl(2)** ).

In the same way polarity is derived for the remaining amino acids. The existence of aromatic side chains dominates two subrepresentations and other properties like the size of amino acids as well restricts to certain representations. In this way amino acid families from representations of semi-simple subalgebras in the chain of symmetry breaking take over the task of physico-chemical main groups. It is instructive to visualize the amino acid families which exist in the second branching step which produces 10 subspaces in the weight diagram. This has been done in picture 3 for the hydrophobic (1) – (1,1) representation of **sp(2)** ⊕ **so(5)** and in picture 4 for the remaining amino acid families.

## IV. POLARITY AND MOLECULAR VOLUME FROM HIGHEST WEIGHTS

So far our inquiry of a mathematical symmetry in the genetic code has lead to qualitative results only. For concluding our work we want to test our findings against hard numerical facts: We try to find expressions for special operators in the universal envelope of the Lie superalgbera **osp(5|2)** which may be interpreted as physical or chemical properties of the encoded amino acids. In a similar way the Hamiltonian of a dynamical system describes its energy levels. In order to bear the correct symmetry these operators have to be invariant against actions of the generators of the Lie superalgebra: Therefore they have to be built from Casimir operators of **osp(5|2)** and the chain of subalgebras which has been considered in this paper. Since any algebraic expression formed with Casimir operators is admissible, it is still a formidable task to find the correct expression for the different properties of amino acids. To reduce this enormous number of possibilities we admit only expressions which are polynomial up to third order in the Casimir operators. From all these expressions we then choose those, which by a minimum of free parameters allow to interpolate a property of the amino acids with a mathematical fitting algorithm.

At first we have to decide which properties we would like to match. Already in the initializing paper on this type of models Hornos and Hornos suggested to examine the properties which have been inquired in Granthams work: polarity and molecular volume. Those have the biggest influence on the usage of amino acids in bio molecules.

A comment seems in order: It is certainly questionable whether this technique can produce significant insight into the mechanism of a complex system. Should the system not be described by examining the fundamental interactions of its elementary particles? However our approach does not adopt an atomistic but rather a holistic approach: After giving the right expression for certain physico-chemical properties we might find the interactions which justify the construction of the operator. If not we still give a useful technique to calculate the properties from a few labels (heigest weight labels) belonging to an irreducible representation.

In the chain of subalgebras **of osp(5|2)** only **so(5)** needs further explanation. (The values for the Casimir - operator $L^2$ in **sl(2)** is given by l (l+2) / 4 as stated in any textbook on Lie algebras.) For the Casimir operators of **so(5)** we cite the results from the work of Wybourne[24] and Kemmer et al.[25]. Instead of using highest weights or Dynkin labels, the eigenvalues of the Casimir invariants can be expressed in a Gelfan'd Zetlin basis, which indicates the number of boxes of the corresponding Young tableau representation. The Young tableaux for the occurring representations of **so(5)** have been given by Forger and Sachse[10] (p.5432). For instance for the representation with highest weight (0,1) the Young tableau has the labels

$$(1/2,\ 1/2) \cong \begin{array}{|c|}\hline S \\ \hline S \\ \hline \end{array} \ ,$$

where s indicates a spinor representation. We summarize the three representations of **so(5)** showing up in the codon representation in a diagram:

| Dynkin label | Young tableau |
|---|---|
| ( 0, 1) | (1/2, 1/2) ≅ [S / S] |
| ( 1, 1) | (3/2, 1/2) ≅ [S □ / S] |
| ( 0, 3) | (3/2, 3/2) ≅ [S □ / S □] |

Table 7: Representations of **so(5)** occurring in the branching scheme of the codon representation of **osp(5|2).**

In general the correspondence between Dynkin labels and Young tableau labels for **so( 2n+1)** – algebras is:

$$D(m_1, m_2, \dots, m_n) \cong Y(f_1, f_2, \dots, f_n), \tag{2}$$

where

$$f_j = m_j + m_{j+1} + \dots + m_{n-1} + \tfrac{1}{2}\, m_n,$$
$$f_n = \tfrac{1}{2}\, m_n,$$

and $j \in \{1, \dots, n\}$ . In our case we have n = 2.

The eigenvalues of the Casimir invariants of **so(5)** have been given by Wybourne[24] in terms of $f_1 = k$ and $f_2 = l$:

$$I_2 = \frac{1}{2}\left(k(k+3) + l(l+1)\right), \qquad I_4 = (k+1)(k+2)\, l\,(l+1). \tag{3}$$

Table 8 summarizes the values of the Casimir Invariants calculated from the weight vectors given in the branching scheme of table 4 and 5.

For interpolating polarity and molecular volume we use the best-fit algorithm of Mathematica[26]. We checked several polynomial functions up to 3rd order in the Casimir invariants of table 8. Here we just state the functions which admitted the best fit for polarity and molar volume. Polarity has been fitted by the function:

$$\begin{aligned} P = {} & \lambda_1^P\, \mathrm{I}_2(so(5)) + \lambda_2^P\, \mathrm{I}_4(so(5)) + \alpha_1^P L_1^2 + \alpha_2^P L_2^2 + \alpha_3^P L_3^2 \\ & + \alpha_{12}^P (L_1 + L_2)^2 + \beta_1^P\, |L_{3,z}| + \beta_2^P\, L_{3,z}^2 \\ & + \gamma_{12}^P\, ((L_1 + L_2)^2 - 2)\, (L_{1,z} + L_{2,z})^2, \end{aligned} \tag{4}$$

with the parameters: $\lambda_1^P = 0{,}5471$; $\lambda_2^P = 0{,}1034$; $\alpha_1^P = 3{,}618$; $\alpha_2^P = 1{,}098$; $\alpha_3^P = -0{,}324$; $\alpha_{12}^P = -0{,}369$; $\beta_1^P = 1{,}518$; $\beta_2^P = -0{,}816$; $\gamma_{12}^P = -0{,}1352$.

In a similar way we find for the molecular volume the function:

$$M = \lambda_1^M \, I_2\big(so(5)\big) + \alpha_1^M L_1^2 + \alpha_2^M L_2^2 + \eta_2^M L_2^4 + \alpha_3^M L_3^2 + \alpha_{12}^M (L_1+L_2)^2 + \eta_{12}^M (L_1+L_2)^4 + \begin{bmatrix} \left(\beta_1^M |L_{3,z}| + \beta_2^M L_{3,z}^2\right) \\ +\gamma_{12}^M\left((L_1+L_2)^2 - 2\right)(L_{1,z}+L_{2,z})^2 \end{bmatrix} (L_2^2 - 3{,}75), \quad (5)$$

with the parameters:
$\lambda_1^M = 37{,}51$; $\alpha_1^M = 32{,}61$; $\alpha_2^M = -46{,}36$; $\eta_2^M = 2{,}347$; $\alpha_3^M = -24{,}23$; $\alpha_{12}^M = 36{,}218$; $\eta_{12}^M = -7{,}79$;
$\beta_1^M = 46{,}74$; $\beta_2^M = -0{,}875$; $\gamma_{12}^M = -2{,}918$.

With these formulas and parameters we find for the interpolation of polarity and molecular volume the results given in table 9.

For both interpolations the mean deviation indicates a reasonable approximation between calculated and experimental values. Such deviations from theoretical results show up for the models described in the first chapter frequently (e.g. energy levels of atoms of high atomic weight, Okubo mass formula, etc...). Modifying the structure of the interpolating functions the approximation can certainly be improved. Here we have presented the most simple functions, only.

Comparing polarity with the property of amino acid families as introduced in table 6, the question arises, why histidine is located in a hydrophobic aa-family. We like to recall that histidine is rather used in protein functional sites due to its ability of binding a proton, which makes it suitable for catalytic processes. In this respect its positive charge doesn't imply that it is hydrophilic. A comparable argument can be found for threonine, which could be considered as being rather hydrophobic than the opposite as in our classification. It should be however recalled, that among "similar" small amino acids threonine rather substitutes with serine which is polar than with alanine which is rather hydrophobic. Evidently our classification underlines this tendency, since the calculated polarity of histidine is smaller than its experimental polarity and for threonine the calculated polarity is slightly higher. Regarding molecular volume we would like to add, that Grantham takes for proline a much lower than its formal value, since the carbon atom is attached to the back bone. This also agrees with our theoretical value.

# V. CONCLUSION AND OUTLOOK

In this article a natural classification scheme for the amino acids of the genetic code has been derived from mathematical symmetries which had been found elsewhere. The classification is natural, as it shows very good agreement with biophysical properties of the amino acids and fits as well into universal features taken from protein evolution and alignment analysis. Therefore our classification scheme is a candidate for explaining the structure of the genetic language. This classification may help to make predictions on protein structure and synthesis. Attaching the right labels (i.e. sets of highest weights or quantum numbers) to amino acids is supposed to make protein research more accessible to numeric methods. For instance we suggest to determine score matrices for protein alignments in phylogenetic studies from these labels which would reduce significantly the degrees of freedom of such an inquiry (i.e. from the dimension of the score matrix – 190 – to the number of the highest weights used to describe the codon representation.)

**ACKNOWLEDGEMENTS**

One of the authors (S.Sachse) would like to thank Dr. Reinhard Schneider from EMBL, Heidelberg for fruitful discussions and pointing out relevant literature on the subject, especially [20] and [21] and Karl-Ludwig Konz from CSC Deutschland for his support in compilation of the programs supplied for the AIS method[19].

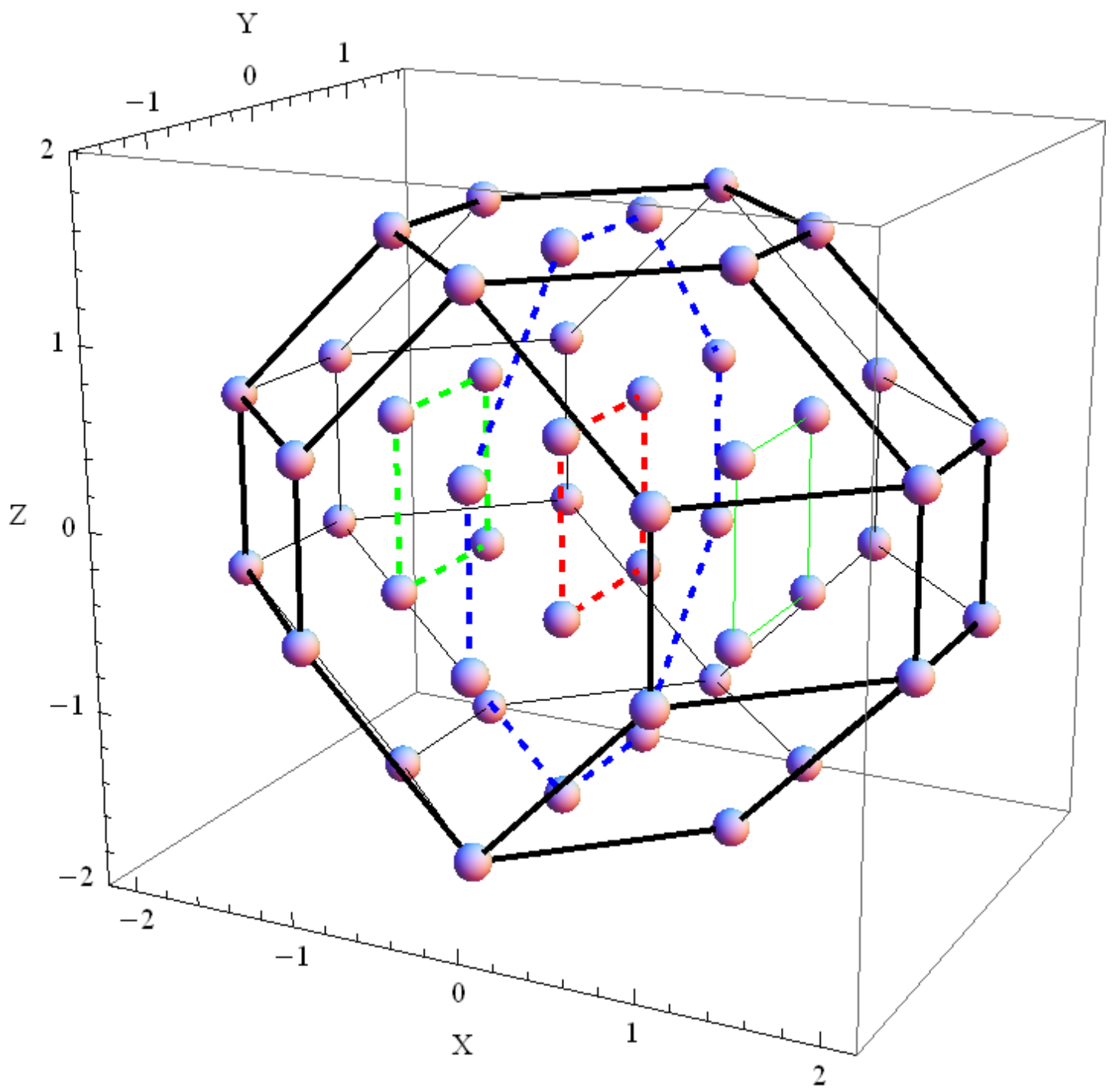

Picture 2: Weight diagram of the codon representation (5/2,0,1) of **osp (5|2).** The weight vectors connected with black and blue lines have multiplicity 1, those connected with green lines have multiplicity 2 and those on the red square have multiplicity 3.

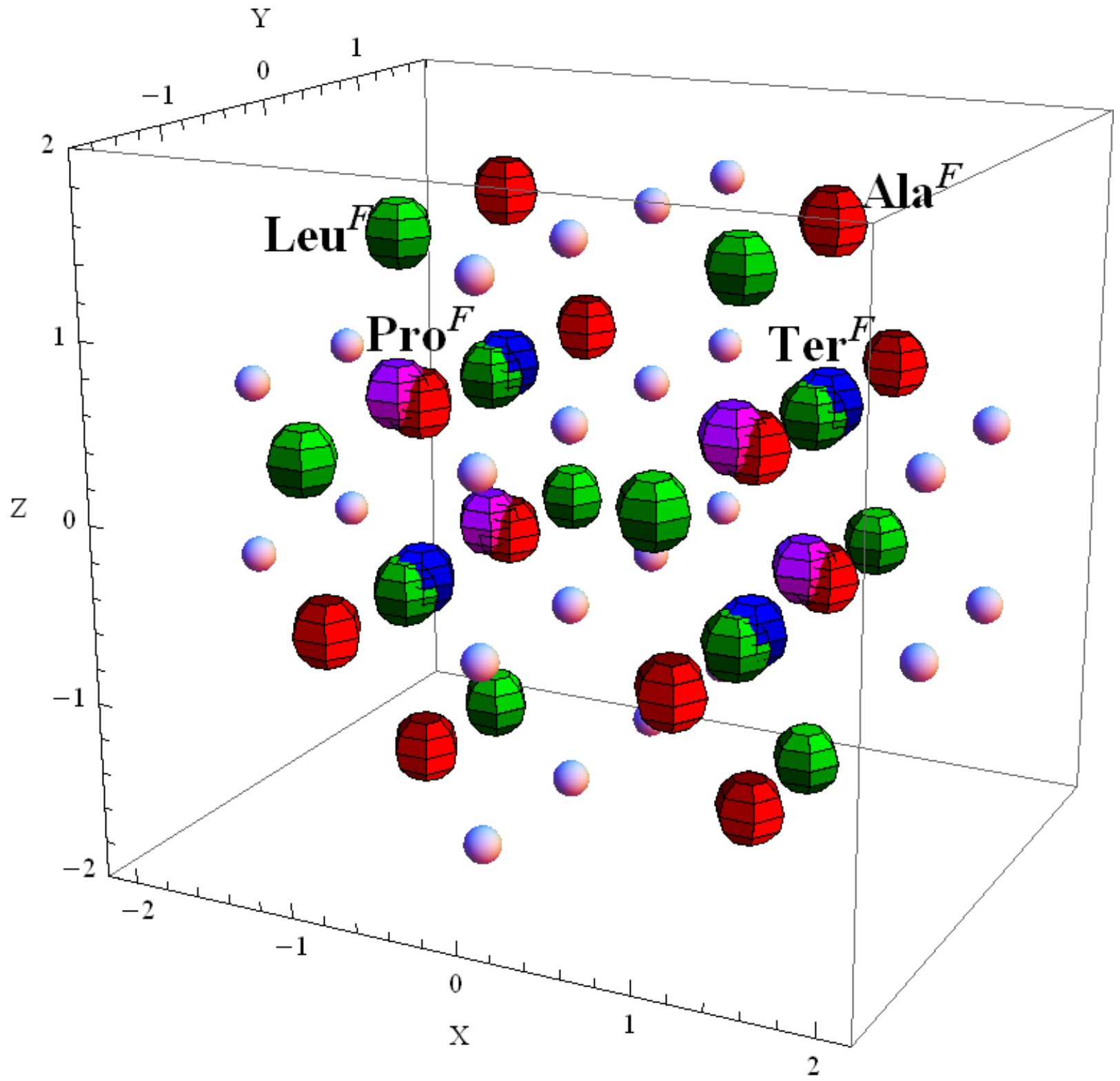

Picture 3: Weight diagram of the codon representation of **osp(5|2).** The colored spheres show amino acid families of the hydrophobic (1) – (1,1) representation of **sp(2) ⊕ so(5).**

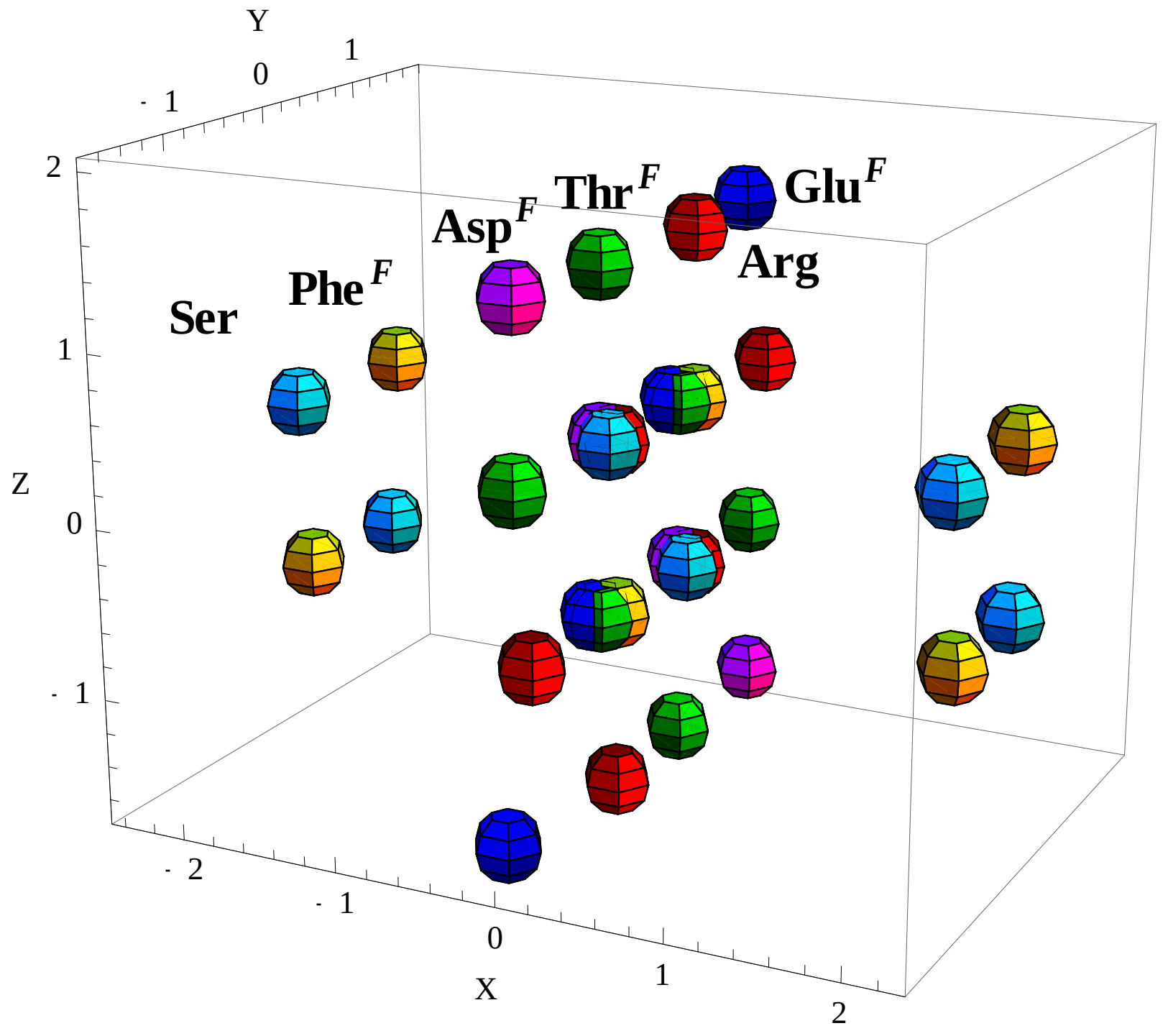

Picture 4: Weight diagram of the codon representation of **osp(5|2)**. The colored spheres show amino acid families of the (0) – (0,3) and (2) – (0,1) h.w. representation of **sp(2)** ⊕ **so(5)**.

| AA | $I_2$ | $I_4$ | $L_1^2$ | $L_2^2$ | $L_3^2$ | $L_{12}^2$ | $\|L_{3,z}\| = m_3$ | $\|L_{12,z}\| = m_{12}$ |
|---|---|---|---|---|---|---|---|---|
| Val | $3\frac{3}{4}$ | $13\frac{1}{8}$ | $\frac{3}{4}$ | 2 | $\frac{3}{4}$ | $3\frac{3}{4}$ | 1 | 3 |
| Ala | $3\frac{3}{4}$ | $13\frac{1}{8}$ | $\frac{3}{4}$ | 2 | $\frac{3}{4}$ | $3\frac{3}{4}$ | 1 | 1 |
| Gly | $3\frac{3}{4}$ | $13\frac{1}{8}$ | $\frac{3}{4}$ | 2 | $\frac{3}{4}$ | $\frac{3}{4}$ | 1 | 1 |
| Leu | $3\frac{3}{4}$ | $13\frac{1}{8}$ | $\frac{3}{4}$ | $\frac{3}{4}$ | 2 | 2 | 2 | 2 |
| Ile | $3\frac{3}{4}$ | $13\frac{1}{8}$ | $\frac{3}{4}$ | $\frac{3}{4}$ | 2 | 2 | 0 | 2 |
| Cys | $3\frac{3}{4}$ | $13\frac{1}{8}$ | $\frac{3}{4}$ | $\frac{3}{4}$ | 2 | 0 | 2 | 0 |
| Met | $3\frac{3}{4}$ | $13\frac{1}{8}$ | $\frac{3}{4}$ | $\frac{3}{4}$ | 2 | 0 | 0 | 0 |
| Ter | $3\frac{3}{4}$ | $13\frac{1}{8}$ | $\frac{3}{4}$ | $\frac{3}{4}$ | 0 | 2 | 0 | 2 |
| Trp | $3\frac{3}{4}$ | $13\frac{1}{8}$ | $\frac{3}{4}$ | $\frac{3}{4}$ | 0 | 2 | 0 | 0 |
| Pro | $3\frac{3}{4}$ | $13\frac{1}{8}$ | $\frac{3}{4}$ | $\frac{3}{4}$ | $\frac{3}{4}$ | $\frac{3}{4}$ | 1 | 1 |
| Arg | $5\frac{1}{4}$ | $65\frac{5}{8}$ | 0 | 2 | $\frac{3}{4}$ | 2 | 1 | 2 |
| Thr | $5\frac{1}{4}$ | $65\frac{5}{8}$ | 0 | $\frac{3}{4}$ | 2 | $\frac{3}{4}$ | 2 | 1 |
| Lys | $5\frac{1}{4}$ | $65\frac{5}{8}$ | 0 | $\frac{3}{4}$ | 2 | $\frac{3}{4}$ | 0 | 1 |
| Gln | $5\frac{1}{4}$ | $65\frac{5}{8}$ | 0 | $3\frac{3}{4}$ | 0 | $3\frac{3}{4}$ | 0 | 3 |
| Glu | $5\frac{1}{4}$ | $65\frac{5}{8}$ | 0 | $3\frac{3}{4}$ | 0 | $3\frac{3}{4}$ | 0 | 1 |
| Asn | $5\frac{1}{4}$ | $65\frac{5}{8}$ | 0 | $3\frac{3}{4}$ | 0 | 0 | 3 | 0 |
| Asp | $5\frac{1}{4}$ | $65\frac{5}{8}$ | 0 | $3\frac{3}{4}$ | 0 | 0 | 1 | 0 |
| Phe | $1\frac{1}{4}$ | $5\frac{5}{8}$ | 2 | $\frac{3}{4}$ | 0 | $3\frac{3}{4}$ | 0 | 3 |
| Tyr | $1\frac{1}{4}$ | $5\frac{5}{8}$ | 2 | $\frac{3}{4}$ | 0 | $3\frac{3}{4}$ | 0 | 1 |
| His | $1\frac{1}{4}$ | $5\frac{5}{8}$ | 2 | $\frac{3}{4}$ | 0 | $\frac{3}{4}$ | 0 | 1 |
| Ser | $1\frac{1}{4}$ | $5\frac{5}{8}$ | 2 | 0 | $\frac{3}{4}$ | 2 | 1 | 2 |

Table 8: Eigenvalues of Casimir invariants in the branching scheme of **osp(5|2)**

| AA | Polarity Lit. | Polarity Calc. | Mol. Vol. Lit. | Mol. Vol. Calc. |
|---|---|---|---|---|
| Val | 5,9 | 5,26 | 84 | 70,6 |
| Ala | 8,1 | 7,16 | 31 | 52,7 |
| Gly | 9 | 8,67 | 3 | 33,9 |
| Leu | 4,9 | 5,33 | 111 | 88,6 |
| Ile | 5,2 | 5,56 | 111 | 124,5 |
| Cys | 5,5 | 6,07 | 55 | 47,4 |
| Met | 5,7 | 6,30 | 105 | 83,2 |
| Trp | 5,4 | 6,21 | 170 | 173,0 |
| Pro | 8 | 7,30 | **55** | 46,3 |
| Arg | 10,5 | 11,57 | 124 | 90,6 |
| Thr | 8,6 | 9,49 | 61 | 91,0 |
| Lys | 11,3 | 9,72 | 119 | 126,9 |
| Gln | 10,5 | 10,26 | 85 | 82,4 |
| Glu | 12,3 | 12,15 | 83 | 82,4 |
| Asn | 11,6 | 10,98 | 56 | 56,1 |
| Asp | 13 | 14,48 | 54 | 56,1 |
| Phe | 5,2 | 5,81 | 132 | 150,9 |
| Tyr | 6,2 | 7,70 | 136 | 120,3 |
| His | 10,4 | 9,21 | 96 | 90,5 |
| Ser | 9,2 | 8,22 | 32 | 36,4 |
| rms | | 0,88 | | 16,8 |

Table 9: Comparision of measured and calculated values for polarity and molecular volume, rms: root mean square for the deviation of theoretical values from measurement.

## REFERENCES

[1] M. Gell-Mann and Y. Ne'eman: *The eightfold way,* Benjamin, New York (1964).
[2] H. Georgi and S.L. Glashow, Phys. Rev. Lett. **32**, 438.
[3] An introduction to symmetry principles in high energy physics is given in:
T-P. Cheng and L-F. Li: *Gauge theories of elementary particle physics*, Oxford University Press (1984). We refer to the references listed therein.
[4] J.E.M. Hornos, Y.M.M. Hornos and M. Forger*: Symmetry and Symmetry Breaking: An Algebraic Approach to the Genetic Code*, Phys. Rev. **E 56** (1999) 7078-7182.
[5] F. Iachello, Chem. Phys. Lett. **78** (1981) 581, A. Arima and F. Iachello, Ann. Phys. **99** (1976) 253,
[6] J.E.M. Hornos and Y.M.M. Hornos*, Algebraic Model for the Evolution of the Genetic Code*, Phys. Rev. Lett. **71,** 4401-4404 (1993).
[7] J. Maddox: *The Genetic Code by Numbers*, Nature **367**, 111 (1994).
[8] I. Stewart: *Broken Symmetry in the Genetic Code?*, New Scientist **141**, No. 1915, 16 (1994).
[9] M. Forger, S. Sachse: *Lie Superalgebras and the Multiplet Structure of the Genetic Code I: Codon Representations*, J. Math. Phys**. 41** (2000) 5407-5422.
[10] M. Forger, S. Sachse *Lie Superalgebras and the Multiplet Structure of the Genetic Code II: Branching Rules,* J. Math. Phys. **41** (2000) 5423-5444.
[11] J.D. Bashford, I. Tsohantjis and P.D. Jarvis*, Codon and Nucleotide Assignments in a Supersymmetric Model of the Genetic Code*, Phys. Lett. **A 233** (1997) 481-488.
and J.D. Bashford, I. Tsohantjis and P.D. Jarvis, *A Supersymmetric Model for the Evolution of the Genetic Code*, Proc. Nat. Acad. Sci. USA **95** (1998) 987-992.
[12] Robert B. Russell, Matthew J. Betts & Michael R. Barnes: *Amino acid properties*, http://www.russelllab.org/aas/.
[13] http://en.wikipedia.org/wiki/Genetic_Code.
[14] http://en.wikipedia.org/wiki/Cysteine.
[15] M. Dayhoff: *Atlas of Protein Sequence and Structure*, Natl. Biomed. Res. Found., Washington **5**, Suppl. 3 (1978).
[16] S. Henikoff, J.G. Henikoff: *Amino Acid Substitution Matrices from Protein Blocks*, Proc. Natl. Acad. Sci. USA, **89** (1992) 10915-10919.
[17] G.H. Gonnet, R. Scholl: *Scientific Computation,* Cambridge University Press (2009).
[18] C. Kosiol et al.: *A new criterion and method for amino acid classification,* Journal of Theoretical Biology **228** (2004) 97-106.
and C. Kosiol: *Markov models for protein sequence evolution*, PhD Thesis. Cambridge University (2006).
[19] http://www.ebi.ac.uk/goldman/AIS/
[20] A.C.W. May*: Towards more meaningful hierarchical classification of amino acid scoring matrices*, Protein Engineering **12** (1999) 707-712.
[21] T.D. Wu, D.L. Brutlag: *Discovering empirically conserved amino acid substitution groups in databases of protein families*, Proc Int Conf Intell Syst Mol Biol. **4** (1996) 230-40.
[22] R. Schneider and C. Sander: *The HSSP database of protein structure-sequence alignments*, Nucleic Acids Research, **24** (1996) 201–205.
[23] F. Aantoneli, M. Forger, P.A. Gaviria and J.E.M. Hornos: *On amino acid and codon assignment in algebraic models for the genetic code,* International Journal of Modern Physics B **24** (2010) 435-463.
[24] B.G. Wybourne: *Symmetry principles in atomic spectroscopy,* Wiley, New York (1970).
[25] N.Kemmer, D.L. Pursey and S.A. Williams: *Irreducible representations of the five dimensional rotation group I*, J. Math. Phys. **9** (1968) 1224–1230.
[26] The functions for polarity and for molecular volume have been fitted using **Wolfram** *Mathematica* 8.0.1.